\documentclass[conference]{IEEEtran}
\ifCLASSINFOpdf
\else
\fi
\usepackage{amsmath,dsfont,bbm,epsfig,amssymb,amsfonts,amstext,verbatim,amsopn,cite,subfigure,multirow,multicol,lipsum,xfrac}
\usepackage{amsthm}
\usepackage{mathtools,amsthm}
\usepackage{perpage}
\usepackage{balance}
\usepackage{url}
\usepackage{amsfonts}
\usepackage{epsfig}
\usepackage[font={small}]{caption}
\usepackage{psfrag}	
\usepackage{etoolbox}
\usepackage{algorithmicx}
\usepackage[Algorithm,ruled]{algorithm}
\usepackage{algpseudocode}
\usepackage{pifont}
\usepackage[utf8]{inputenc}
\usepackage[T1]{fontenc}  
\usepackage[nolist]{acronym}
\MakePerPage{footnote}
\usepackage{paralist}
\usepackage{enumitem}
\usepackage{bbm}
\usepackage[process=auto]{pstool}
\usepackage{tikz}
\usetikzlibrary{shapes,arrows}

\newcommand{\setA}{\mathbbmss{A}}

\newcommand{\setE}{\mathbbmss{E}}
\newcommand{\setR}{\mathbbmss{R}}

\newcommand{\setS}{\mathbbmss{S}}

\newcommand{\setB}{\mathbbmss{B}}

\newcommand{\setC}{\mathbbmss{C}}

\newcommand{\her}{\mathsf{H}}

\newcommand{\mar}{\mathcal{R}}
\newcommand{\mae}{\mathcal{E}}

\newcommand{\mas}{\mathcal{S}}

\newcommand{\mam}{\mathcal{M}}

\newcommand{\bh}{{\mathbf{h}}}

\newcommand{\bx}{{\boldsymbol{x}}}

\newcommand{\set}[1]{\left\lbrace#1\right\rbrace}

\newcommand{\bz}{{\boldsymbol{z}}}
\newcommand{\ba}{{\mathbf{a}}}
\newcommand{\bd}{{\mathbf{d}}}
\newcommand{\bg}{{\mathbf{g}}}
\newcommand{\bs}{{\boldsymbol{s}}}
\newcommand{\br}{{\mathbf{r}}}

\newcommand{\bw}{{\boldsymbol{w}}}

\newcommand{\bb}{{\mathbf{b}}}

\newcommand{\by}{{\boldsymbol{y}}}

\newcommand{\trp}{\mathsf{T}}

\newcommand{\mA}{\mathbf{A}}
\newcommand{\mE}{\mathbf{E}}

\newcommand{\mI}{\mathbf{I}}

\newcommand{\mJ}{\mathbf{J}}
\newcommand{\mG}{\mathbf{G}}

\newcommand{\mD}{\mathbf{D}}

\newcommand{\mH}{\mathbf{H}}

\newcommand{\cir}{\mathrm{cr}}
\newcommand{\sync}{\mathrm{sync}}

\newcommand{\avg}{\mathrm{avg}}

\newcommand{\E}{\mathbb{E}\hspace{.5mm}}

\newcommand{\mrt}{{\mathsf{mrt}}}
\newcommand{\zf}{{\mathsf{zf}}}
\newcommand{\rzf}{{\mathsf{rzf}}}
\newcommand{\Pre}{{\mathrm{Pre}}}

\newcommand{\sinr}{{\mathrm{SINR}}}

\newcommand{\argmax}{\mathop{\mathrm{argmax}}}

\newcommand{\norm}[1]{\lVert #1 \rVert}
\newcommand{\re}[1]{\mathsf{Re}\left\lbrace #1 \right\rbrace}

\newcommand{\abs}[1]{\lvert #1 \rvert}
\newcommand{\tr}[1]{\mathrm{Tr} \{ #1 \}}

\newtheoremstyle{mystyle}
  {}
  {}
  {}
  {}
  {\bfseries}
  {:}
  { }
  {}

\theoremstyle{mystyle}

\algnewcommand\algorithmicLet{\textbf{Let}}
\algnewcommand\Let{\item[\algorithmicLet]}
\algnewcommand\algorithmicSet{\textbf{Set}}
\algnewcommand\Set{\item[\algorithmicSet]}

\algnewcommand\algorithmicInitiate{\textbf{Initiate}}
\algnewcommand\Initiate{\item[\algorithmicInitiate]}
\algnewcommand\algorithmicStart{\textbf{Begin}}
\algnewcommand\Begin{\item[\algorithmicStart]}
\algnewcommand\algorithmicEnd{\textbf{End}}
\algnewcommand\End{\item[\algorithmicEnd]}

\algnewcommand\algorithmicOutP{\textbf{Output:}}
\algnewcommand\Out{\item[\algorithmicOutP]}

\algnewcommand\algorithmicInP{\textbf{Input:}}
\algnewcommand\In{\item[\algorithmicInP]}
\newcommand\NoDo{\renewcommand\algorithmicdo{}}

\newcounter{bar}

\begin{document}
\title{Stepwise Transmit Antenna Selection in Downlink Massive Multiuser MIMO}

\author{
\IEEEauthorblockN{
Ali Bereyhi, 
Saba Asaad, 
Ralf R. M\"uller
}
\IEEEauthorblockA{
Friedrich-Alexander Universit\"at Erlangen-N\"urnberg (FAU)\\
ali.bereyhi@fau.de, saba.asaad@fau.de, ralf.r.mueller@fau.de
\thanks{This work was supported by the German Research Foundation, Deutsche Forschungsgemeinschaft (DFG), under Grant No. MU 3735/2-1.}
}
}


\IEEEoverridecommandlockouts

\maketitle

\begin{abstract}
Due to the large power consumption in RF-circuitry of massive MIMO systems, practically relevant performance~measures such as energy efficiency or bandwidth efficiency are neither necessarily monotonous functions of the total transmit power~nor the number of active antennas. Optimal antenna~selection~is~how- ever computationally infeasible in these systems. In this paper, we propose an iterative algorithm to optimize the transmit power and the subset of selected antennas subject~to~non-monotonous performance measures in massive multiuser MIMO settings.~Nu- merical results are given for energy efficiency and demonstrate that for several settings the optimal number of selected antennas reported by the proposed algorithm is significantly smaller than the total number of transmit antennas. This fact indicates that antenna selection in several massive MIMO scenarios not only reduces the hardware complexity and RF-costs, but also enhances the energy efficiency of the system.
\end{abstract}

\IEEEpeerreviewmaketitle
%

\section{Introduction}
\label{sec:intro}
The ever increasing growth in the number of digital devices such as smart phones, tablets and personal computers as well as broadband wireless applications put a heavy strain~on~wireless community to design the next generation~of~cellular~networks. Considering this backdrop, wireless technologies are shifting towards systems with higher transmission rates and lower power consumption; or more precisely systems with high power efficiency \cite{verdu2002spectral}. In this respect, massive \ac{mimo} has been~considered~as~one~of leading candidates for future wireless communications \cite{larsson2014massive,hoydis2013massive}. 

The so-called ``power scaling law'' in massive \ac{mimo} systems demonstrates that for a given transmission rate, the required transmit power is effectively scaled. More precisely, for a \ac{bs} with a large transmit antenna array, the transmit power of each user is reduced substantially proportional to the number of antennas \cite{ngo2013energy}. Despite the power scaling law, the large antenna arrays employed in massive \ac{mimo} systems increase the consumed power at the \ac{rf}-chains. In other words, although in massive \ac{mimo} systems the radiated power per each transmit antenna is extremely low, the growth in the number of antennas increases the power consumed in the \ac{rf} circuits which is in general linearly scaled with the array size \cite{cui2004energy}. In addition to the large power consumption, massive \ac{mimo} systems with a dedicated \ac{rf}-chain at each antenna suffer from multiple other issues such as high implementation cost and hardware complexity. Considering this fact, it is concluded that despite the appealing theoretical benefits of massive \ac{mimo}, it still needs to be developed from the implementational viewpoints. Recent studies have addressed some of these issues by investigating approaches such as antenna selection \cite{molisch2005capacity}, hybrid analog and digital precoding schemes \cite{liang2014low} and spatial modulation \cite{di2014spatial}, in the context of massive \ac{mimo}. 

\hspace*{-1mm}In this study, we focus on antenna selection which we believe is a promising approach for cost and complexity alleviation in massive \ac{mimo} settings. In general, the optimal antenna selection approach is computationally infeasible in these systems, since it requires an exhaustive search. Consequently, studies in this direction either investigate the asymptotic performance of optimal approaches by means of advanced mathematical tools such as the replica method \cite{bereyhi2017nonlinear,bereyhi2017asymptotics,sedaghat2017new,sedaghat2018least}, or develop and analyze suboptimal algorithms with feasible complexity \cite{asaad2017tas}. A body of recent work has demonstrated that by using effective~antenna selection strategies, several asymptotic properties of massive \ac{mimo} systems, such as energy efficiency \cite{wang2017antenna} and channel hardening \cite{asaad2017tas}, are essentially maintained with significantly fewer \ac{rf} chains. Antenna selection has been moreover shown to be enhancing in several scenarios. For example, the study in \cite{asaad2017optimal} demonstrated that the secrecy performance of a wiretap channel with large antenna arrays can be significantly improved via antenna selection compared to the case with all antennas being active. Accounting for power consumption in \ac{rf}-circuitry, energy efficiency, bandwidth efficiency, and other practically relevant performance measures are neither necessarily monotonous functions of the total transmit power nor the number of active antennas. This fact depicts that antenna selection, as well as transmit power control in general not only can reduce the hardware cost, but also may enhance the performance of the system. In this paper, we propose an iterative method to optimize the total transmit power and the subset of selected antennas for a desired performance measure.

From mathematical points of view, optimal antenna selection is equivalent to the problem of finding an efficient subset of integers which maximizes an objective function subject to a set of constraints. Such problems also arise in other contexts such as pattern recognition and data mining \cite{duda2012pattern,han2011data}. There are therefore a variety of iterative approaches which determine a suboptimal, but effective, subset with low computational complexity among which stepwise regression methods have obtained more credit \cite{duda2012pattern}. Based on these greedy methods, several studies invoked the results in the literature to develop iterative antenna selection algorithms for conventional single-user \ac{mimo} systems\footnote{By conventional \ac{mimo} systems we mean systems with few number of antennas at the receive and transmit sides.} \cite{gorokhov2003receive, gharavi2004fast, sanayei2004capacity,zhou2014iterative} considering the spectral efficiency as the performance measure. Regarding the promising results in the literature of conventional \ac{mimo} and considering the large nature of massive \ac{mimo} systems,~the~stepwise~reg- ression methods can propose effective selection algorithms in these systems with feasible computational complexity.

\subsection*{Contributions}
In this paper, we consider the downlink massive multiuser \ac{mimo} scenario with linear precoding and develop a class of iterative antenna selection algorithms based on stepwise regression methods to optimize the total transmit power and the subset of selected antennas subject to a non-monotonous performance measure. Given a set of constraints on the number of possible active antennas and transmit power, these algorithms iteratively select a subset of antennas and tune the transmit power such that the desired performance measure is maximized. In addition to significant gains reported by the proposed algorithm, numerical results are given for energy efficiency and demonstrate that the antenna selection can also enhance the energy efficiency compared to the case of full antenna selection in some massive \ac{mimo} settings.
%
%
%
%
%
%
%
%
%
%
%
%
\subsection*{Notation}
Throughout the paper the following notations are adopted. Scalars, vectors and matrices are represented with non-bold, bold lower case and bold upper case letters, respectively.~$\mH^{\her}$, $\mH^{\trp}$ and $\mH^{*}$ indicate the Hermitian, transpose and conjugate of $\mH$, respectively, and $\mI_N$ is the $N\times N$ identity matrix. The Euclidean norm of~$\bx$ is denoted by $\norm{\bx}$ and $\log\left(\cdot\right)$ indicates the~binary logarithm. Moreover, $\setE\set{\cdot}$ represents the mathematical expectation. The subset of $\setA$ containing the members of $\setA$ which do not belong to $\setB$ is moreover denoted by $\setA\backslash \setB$. For the sake of compactness, we abbreviate the set of integers $\set{1, \ldots, N}$ with $[N]$.
\section{Problem Formulation}
\label{sec:sys}
Consider a downlink transmission in a multiuser \ac{mimo} system with $K$ single-antenna users. The base station is equipped with a transmit antenna array of length $N$ and employs the \ac{tas} protocol $\mas$ to select a subset of $L$ antennas for transmission. Let $y_k \in \setC$ denote the symbol received at user $k$ for $k\in[K]$ and $\mG \in \setC^{N\times K}$ represent the matrix of channel coefficients between user terminals and the transmit antenna array. It is assumed that the system operates in standard \ac{tdd} mode meaning that the channel is reciprocal in the uplink and doswnlink. In this case, by using the reciprocity of the channel, the received vector $\by \coloneqq [y_1, \ldots, y_K]^\trp$ is written as
\begin{align}
\by= \mH^\trp_L \bx + \bz
\end{align}
where $\bx\in\setC^{L\times 1}$ denotes the transmit vector and~contains~the symbols being transmitted over $L$ active antennas, $\bz\in\setC^{K\times 1}$ is circularly symmetric zero-mean and unit-variance complex Gaussian noise, and $\mH_L \in \setC^{L\times K}$ describes the~channel~between users and $L$ transmit antennas selected by $\mas$, i.e., $\mH_L=\mas(\mG;L)$. The transmit vector $\bx$ is moreover constructed at the base station by linear precoding over active antennas, i.e.,
\begin{align}
\bx=\sqrt{P} \hspace*{.5mm} \mA_L \bs
\end{align}
where $\bs \coloneqq [s_1, \ldots, s_K]^\trp$ with $s_k\in\setC$ being the~data~symbol for user $k$, and $\mA_L\in\setC^{L\times K}$ represents the precoding matrix for $\mH_L$, i.e., $\mA_L=\Pre (\mH_L)$. It is moreover assumed that $\E \abs{s_k}^2 =1$ and the total transmit power is $P$, or equivalently $\E\tr{ \mA_L\mA_L^\her} = 1$.
\subsection{Achievable Rates and Power Consumption Model}
The achievable rate at user $k$ is bounded from below~by~\cite{caire2010multiuser}
\begin{align}
R_k ( L,P )=\log \left(1+ \sinr_k(L,P) \right) \label{achieve}
\end{align}
with $\sinr_k(L,P)$ being defined for $k\in[K]$ as
\begin{align}
\sinr_k(L,P) \coloneqq \frac{t_k(\mH_L , \mA_L) P}{1+ u_k (\mH_L,\mA_L ) P } \label{SINR}
\end{align}
where $ t_k(\mH_L , \mA_L) \coloneqq \abs{\bh_{k}^\trp (L) \ba_{k} (L) }^2$ with $\bh_{k} (L)$ and $\ba_{k} (L)$ denoting the $k$th column vector of $\mH_L$ and $\mA_L$, respectively, and $u_k (\mH_L,\mA_L )$ is the $k$th multiuser interference coefficient defined as
\begin{align*}
u_k (\mH_L,\mA_L ) \coloneqq \sum_{j=1, j\neq k}^K \abs{\bh_{k}^\trp (L) \ba_{j} (L)}^2.
\end{align*}
The total power consumed in this system, including the receive side, is moreover modeled as \cite{ngo2013energy}
\begin{align}
Q(L,P)= \xi P + L Q^{\rm Tx}_\cir + K Q^{\rm Rx}_\cir + \left( K+1 \right) Q_\sync \label{power_model}
\end{align}
where $\xi$ represents the inverse efficiency factor of the power amplifiers used at the base station, $Q^{\rm Tx}_\cir$ and $Q^{\rm Rx}_\cir$ are respectively the circuit power consumed at each transmit and receive \ac{rf}-chain, and $Q_\sync$ denotes the power consumption at the local oscillators utilized at the base station and user terminals for frequency synthesis.
\subsection{Nonuniform Average Spectral Efficiency}
Let $w_k\in \setR^+$ be a weighting coefficient which~describes~the service quality required for $k$. The average spectral efficiency with respect to the vector $\bw=[w_1,\ldots,w_K]^\trp$ is then given~by
\begin{align}
\mar(L,P|\bw) \coloneqq \frac{1}{K} \sum_{k=1}^K w_k R_k(L,P).
\end{align}
This nonuniform average spectral efficiency recovers the conventional definition of the average spectral efficiency by setting $w_k=1$ for $k\in[K]$. Moreover, nonuniform choices~of~weights address cases in which different users enjoy different priorities.
\subsection{Nonuniform Average Energy Efficiency}
The average energy efficiency for a given $\bw$ is defined as
\begin{align}
\mae(L,P|\bw)\coloneqq \frac{\mar(L,P|\bw)}{Q(L,P)}
\end{align}
which determines the average data rate per each unit of energy consumed in this system.

\section{Iterative \ac{tas} and Power Control Algorithm}
In this setting, we deal with two correlated challenges of \ac{tas} and transmit power control. More precisely, the measure of performance in general does not necessarily grow monotonically in terms of the number of active antennas $L$ and the total transmit power $P$. For several measures, there are some scenarios in which the performance is optimized at some $L<L_{\max}$ and $P<P_{\max}$ where $L_{\max}$ and $P_{\max}$ are the maximum number of selected antennas and maximum possible transmit power constrained by hardware restrictions. An example of such a scenario is the case with the average energy efficiency as a measure of performance. In this case, the increase in the number of active antennas as well as the growth in the total transmit power can increase both the average spectral efficiency and the consumed power, and consequently, the energy efficiency could be maximized at some $L$ and $P$ such that $L<L_{\max}$ and $P<P_{\max}$.

In order to formulate this problem, we denote the measure of performance for a weighting vector $\bw$ with $\mam(\setS_L,L,P|\bw)$ when $L$ transmit antennas, whose indices are in $\setS_L\subseteq [N]$, are selected, and the total transmit power is set to $P$. In this case, the problem of optimal \ac{tas} and transmit power control is to find $L^\star$, the set $\setS^\star_{L^\star}$ and the transmit power $P^\star$ such that
%
\begin{align}
(\setS^\star_{L^\star},L^\star,P^\star) = \argmax_{\substack{\setS_L \subseteq [N], L\leq L_{\max} \vspace*{1mm} \\ 0\leq P\leq P_{\max} }} \ \ \mam(\setS_L,L,P|\bw) \label{eq:Opt_main}
\end{align}
for some $L_{\max}$ and $P_{\max}$. Some examples for $\mam(\setS_L,L,P|\bw)$ are $\mar(L,P|\bw)$ and $\mae(L,P|\bw)$. For the sake of brevity, we further drop the argument $\setS_L$ and represent the performance measure with $\mam(L,P|\bw)$ in the rest of this manuscript.

The optimization problem in \eqref{eq:Opt_main} is computationally infeasible due to the large number of searches needed for \ac{tas}. In fact, even for cases in which $\mam(\setS_L,L,P|\bw)$ is convex with respect to $P$, antenna selection needs an exhaustive search of size $\binom NL$ which grows significantly large in $N$. We therefore take an alternative approach based on the stepwise regression method and design an iterative algorithm which jointly selects transmit antennas and tunes the total transmit power $P$.
\subsection{Primary Assumptions and Notations}
In the following sections, we develop iterative algorithms for a large class of performance measures. For deriving the algorithms, we consider the following set of assumptions.
\begin{itemize}
\item The number of selected transmit antennas is limited by $L_{\max}$, i.e., $L\leq L_{\max}$ antennas are selected.
\item The maximum possible transmit power is $P_{\max}$ which means that $P\leq P_{\max}$ in each transmission interval.
\item The performance measure $\mam(L,P|\bw)$ is linearly written in terms of $R_k(L,P)$, e.g., $\mam(L,P|\bw)=\mar(L,P|\bw)$, $\mam(L,P|\bw)=\mae(L,P|\bw)$ or some linear combination of spectral and energy efficiency.
\end{itemize}

The algorithms start from $\ell=1$ and select antennas up to some $L$ in which $L\leq L_{\max}$. For initialization, we set the antenna with strongest channel to be the first selected antenna which means $\mH_1\in\setC^{1\times K}$ is chosen to be the row of $\mG$ with maximum norm. Consequently, the initial precoding matrix is given by $\mA_1 = \Pre(\mH_1)$. In this case, we set the initial value for the transmit power to be $P_1$ such that
\begin{align}
P_1=\argmax_{0\leq P \leq P_{\max}} \mam (1,P|\bw).
\end{align}
Here, $\mam (1,P|\bw)$ represents the performance measure, e.g., $\mar (1,P|\bw)$ or $\mae (1,P|\bw)$, considering the channel matrix to be $\mH_1$ and the precoding matrix to be $\mA_1$. With~this~initialization, we derive a stepwise update rule for~each~step.

\subsection{Stepwise Update Rule}
Assume that we have already selected $\ell < L_{\max}$ transmit antennas, and now intend to select a new antenna. We represent the index of the selected antenna in step $\ell+1$ with $n^\star$ where $n^\star \in [N] \backslash \setS(\ell)$ with $\setS(\ell)$ containing the indices of the antennas selected in steps $[\ell]$. Let us denote the $n^\star$th row of $\mG$ with the vector $\bg^\trp(n^\star)=[g_1(n^\star), \ldots, g_k(n^\star)]$. Without loss of generality, one can write
\begin{subequations}
\begin{align}
\hspace*{-2mm} \bh_k(\ell+1)&\hspace*{-.7mm}=\hspace*{-.7mm}[\bh_k^\trp (\ell),  \ g_k(n^\star)]^\trp \\
\hspace*{-2mm}\ba_k(\ell+1)&\hspace*{-.7mm}=\hspace*{-.7mm} [\sqrt{\mu(\ell,n^\star)} \hspace*{1mm} \ba_k^\trp(\ell) + \bd_k^\trp (\ell,n^\star), \ b_k(\ell,n^\star)]^\trp \label{update_a}
\end{align}
\end{subequations}
where $\mu(\ell,n^\star)$, $\bd_k (\ell,n^\star)\in\setC^{\ell\times1}$ and $b_k(\ell,n^\star)$ are derived in terms of $\mH_\ell$, $\mA_\ell$ and $\bg(n^\star)$ depending on the precoding rule $\Pre(\cdot)$. We later derive these parameters for different linear precodings, namely \ac{mrt}, \ac{zf} and \ac{rzf}.

By substituting the new channel and precoding coefficients in \eqref{SINR}, we have
\begin{align}
\sinr_k &(\ell+1,P) = \nonumber \\
&\frac{ \mu(\ell,n^\star) t_k(\mH_\ell , \mA_\ell) P + \epsilon_k(\ell, n^\star) P}{1+ \mu(\ell,n^\star) u_k (\mH_\ell,\mA_\ell ) P+ \psi_k(\ell,n^\star)P } \label{sinr:update}
\end{align}
where $\epsilon_k (\ell, n^\star)$ and $\psi_k (\ell,n^\star)$ are given by
\begin{subequations}
\begin{align}
\epsilon_k (\ell, n^\star) &\coloneqq \abs{\delta_{kk} (\ell, n^\star)}^2 \nonumber \\
&+ 2 \re{ \sqrt{\mu(\ell,n^\star)} \hspace*{1mm} \bh_k^\trp(\ell) \ba_k(\ell) \delta_{kk}^* (\ell, n^\star) }\\
\psi_k (\ell,n^\star) &\coloneqq  \sum_{j=1, j\neq k}^K  \abs{\delta_{kj} (\ell, n^\star)}^2 \nonumber \\
&+ 2 \re{ \sqrt{\mu(\ell,n^\star)} \hspace*{1mm} \bh_k^\trp(\ell) \ba_j(\ell) \delta_{kj}^* (\ell, n^\star) }
\end{align}
\end{subequations}
with $\delta_{kj} (\ell, n^\star)$ being defined as
\begin{align}
\delta_{kj} (\ell, n^\star) \coloneqq  \bh_k^\trp (\ell) \bd_j (\ell,n^\star)  + g_k(n^\star) b_j(\ell,n^\star).
\end{align}
Here, the argument $n^\star$ indicates the dependency on the coefficients of the new selected channel. By substituting \eqref{sinr:update} into \eqref{achieve}, one can finally write
\begin{align}
R_k(\ell+1,P)= R_k(\ell,P)+  \log \theta_k (\ell,n^\star,P) \phi_k (\ell,P) \label{eq:step_R}
\end{align}
where $\phi_k (\ell,P)$ is given by
\begin{align}
\phi_k (\ell,P) \coloneqq  \dfrac{1/P + u_k (\mH_\ell,\mA_\ell )}{1/P + u_k (\mH_\ell,\mA_\ell ) + t_k (\mH_\ell,\mA_\ell ) } \label{phi_k}
\end{align}
and $\theta_k (\ell,n^\star,P)$, which depends  on the selected antenna, reads
\begin{align}
\theta_k (\ell, &n^\star ,P) \coloneqq \nonumber \\
&1 + \dfrac{\mu(\ell,n^\star) t_k(\mH_\ell , \mA_\ell) + \epsilon_k (\ell, n^\star)}{1/P + \mu (\ell,n^\star) u_k (\mH_\ell,\mA_\ell )+ \psi_k (\ell,n^\star)} . \label{theta_k}
\end{align}

The stepwise variation of $R_k(\ell,P)$ in \eqref{eq:step_R} lets us to select a new antenna in step $\ell+1$ given the selected antennas in steps $[\ell]$ and the power $P$, such that $R_k(\ell,P)$ has the maximum growth. Although such a selection does not necessarily give the optimal selection, it leads to an efficient algorithm with low complexity. Starting from \eqref{eq:step_R}, the performance measure is updated in each~step~as
\begin{align}
\mam(\ell+1,P|\bw)= \mam(\ell,P|\bw)+ \Theta (\ell,n^\star,P) \label{stepwise_sel}
\end{align}
where $\Theta (\ell,n^\star,P)$ is derived as a function of $\theta_k (\ell,n^\star,P)$ and $\phi_k (\ell,P)$. We refer to \eqref{stepwise_sel} as the stepwise update rule.
\subsection{Iterative Approach for \ac{tas} and Power Control}
Based on the stepwise update rule, we now propose~an~iterative \ac{tas} and power control algorithm. Here, we state the algorithm for a general precoding and performance measure and give detailed derivations for different measures and precoding matrices in the next sections.

The algorithm follows the stepwise regression method. In each step, we select the antenna which results in the maximum possible growth in the performance measure. Although this approach does not necessarily lead to the optimal solution, it enjoys low computational complexity while demonstrating a better performance compared to other low complexity approaches. Using the stepwise regression method, one starts with the initialization $\mH_1$, $\mA_1$ and $P_1$ for $\ell=1$ and chooses the transmit antenna $n^\star$ in step $\ell$ as
\begin{align}
n^\star = \argmax_{n\in[N]\backslash \setS(\ell)}  \Theta (\ell,n,P_\ell).
\end{align}
The channel matrix is then updated as $\mH_{\ell+1}=[\mH_\ell^\trp, \bg_{n^\star}]^\trp$ and the transmit power for the next step is set to be\footnote{It is worth to note that this optimization involves only a single argument $P$, and its dimension does not grow with the system size.}
\begin{align}
P_{\ell+1} = \argmax_{0\leq P\leq P_{\max}} \mam(\ell+1,P|\bw).
\end{align}
In general, the algorithm can select antennas until it reaches $L_{\max}$. Nevertheless, the stepwise update rule may stop growing at some step $\ell<L_{\max}$. We therefore stop the algorithm at step $L^\star \leq L_{\max}$ when either $L^\star=L_{\max}$ or 
\begin{align}
\Theta (L^\star,n^\star,P_{L^\star}) \leq 0.
\end{align}
The outputs of this algorithm are then $L^\star$, $P_{L^\star}$ and $\setS(L^\star)$. The algorithm is illustrated in Algorithm~\ref{alg1} in details.\\

\begin{algorithm}[t]
\caption{Iterative \ac{tas} and Power Control}
\label{alg1}
\begin{algorithmic}[0]
\In $\mG$, $P_{\max}$, $L_{\max}$, precoding rule $\Pre(\cdot)$ and performance measure $\mam(\ell,P|\bw)$
\Initiate Let $\ell=1$ and
\begin{align}
n^\star = \argmax_{n\in[N]} \norm{\mG(n, :)}
\end{align}
and set $\mH_{1} = \mG(n^\star, :)$ and $\mA_1=\Pre(\mH_1)$. Set
\begin{align}
P_1=\argmax_{0\leq P \leq P_{\max}} \mam(1,P| \bw)
\end{align}
and $\setS(1)= \set{n^\star}$. \vspace*{2mm}
\While\NoDo $\ell< L_{\max}$
\begin{align}
n^\star &= \argmax_{n\in[N]\backslash\setS(\ell)} \Theta(\ell,n,P_\ell)
\end{align}
let $\Theta^\star= \Theta(\ell,n^\star,P_\ell)$.\vspace*{1.5mm}
\If{$\Theta^\star \leq 0$}

break
\EndIf \vspace*{1.5mm}\\

Set $\mH_{\ell+1}= \left[\mH_\ell^\trp, \hspace*{1mm} \mG(n^\star, :)^\trp \right]^\trp$ and
\begin{subequations}
\begin{align}
\mD(\ell,n^\star) &= \left[ \bd_1 (\ell,n^\star), \ldots, \bd_K (\ell,n^\star) \right] \\
\bb (\ell,n^\star) &= \left[ b_1 (\ell,n^\star), \ldots, b_K (\ell,n^\star) \right]^\trp,
\end{align}
\end{subequations}
and update the precoding matrix as
\begin{align}
\mA_{\ell+1}= \begin{bmatrix}
         \sqrt{\mu (\ell,n^\star) } \mA_{\ell} +\mD(\ell,n^\star)   \\
         \bb^\trp (\ell,n^\star)
         \end{bmatrix}.
\end{align}
Define the scalar function
\begin{align}
\mam(\ell+1,P|\bw)= \mam(\ell,P|\bw)+ \Theta (\ell,n^\star,P)
\end{align}
and  update the transmit power as
\begin{align}
P_{\ell+1}=\argmax_{0\leq P \leq P_{\max}} \mam(\ell+1,P| \bw).
\end{align}
Update $\setS(\ell+1)= \setS(\ell) \cup \set{n^\star}$ and $\ell=\ell+1$.\vspace*{1mm}
\EndWhile \vspace*{2mm}
\Out $L^\star=\ell$, $P_{L^\star}=P_\ell$ and $\setS(L^\star)=\setS(\ell)$.
\end{algorithmic}
\end{algorithm}

The proposed algorithm considers a general setting. In the sequel, we derive the detailed formulation for several performance measures and linear precoders. Namely we determine the stepwise update rule for the average spectral and energy efficiency while considering \ac{mrt}, \ac{zf} and \ac{rzf} precoders.
\section{Derivations for Different Measures}
As mentioned before, the average spectral efficiency and the average energy efficiency are some common choices for the performance measure. We therefore derive explicitly the stepwise update rule for each of these measures. 
\subsection{Average Spectral Efficiency}
For this performance measure $\mam(\ell,P|\bw)$ is given by
\begin{align}
\mam(\ell,P|\bw) = \mar(\ell,P|\bw) = \frac{1}{K} \sum_{k=1}^K w_k R_k (\ell,P).
\end{align}
Therefore, $\Theta(\ell,n^\star,P)$ in this case reads
\begin{align}
\Theta (\ell,n^\star,P) = \frac{1}{K} \sum_{k=1}^K w_k \log \theta_k (\ell,n^\star,P) \phi_k (\ell,P).
\end{align}
\subsection{Average Energy Efficiency}
In this case, we have
\begin{align}
\mam(\ell,P|\bw) = \mae(\ell,P|\bw)= \frac{\mar(\ell,P|\bw)}{Q(\ell,P)}
\end{align}
with $Q(\ell,P)$ being defined in \eqref{power_model}. Noting that 
\begin{align}
Q(\ell+1,P) = Q(\ell,P) + Q_{\cir}^{\rm Tx}
\end{align}
the stepwise growth term is found as in \eqref{eq:Thet_Big} on the top of the next page.
\begin{figure*}[!t]
\begin{align}
\Theta (\ell,n^\star,P) = \frac{1}{Q(\ell+1,P)} \left[  \frac{1}{K} \sum_{k=1}^K w_k \log \theta_k (\ell,n^\star,P) \phi_k (\ell,P) - Q^{\rm Tx}_{\cir} \mae(\ell,P|\bw) \right] . \label{eq:Thet_Big}
\end{align}
\centering\rule{13cm}{0.01pt}
\begin{align}
\Delta (\ell,n^\star;\lambda) = &\frac{\norm{\br (\ell,n^\star;\lambda)}^2}{1+\bg^\her(n^\star) \mJ_\ell(\lambda)^{-1} \bg(n^\star)} + \br^\her (\ell,n^\star;\lambda) \mE(\ell,n^\star;\lambda) \br (\ell,n^\star;\lambda) \label{eq:Delta_Big}
\end{align}
\centering\rule{17cm}{0.1pt}
\vspace*{2pt}
\end{figure*}

\section{Derivations for Different Linear Precoders}
For different linear precoders, the terms $\mu(\ell,n^\star)$, $\bd_k (\ell,n^\star)$ and $b_k(\ell,n^\star)$ are of different forms. In this section, we derive these terms for the \ac{mrt}, \ac{zf} and \ac{rzf} precoders. For the sake of compactness, we define the matrix $\mD(\ell,n^\star)\in \setC^{\ell\times K}$ and the vector $\bb (\ell,n^\star) \in \setC^{K\times 1}$ as
\begin{subequations}
\begin{align}
\mD(\ell,n^\star) &= \left[ \bd_1 (\ell,n^\star), \ldots, \bd_K (\ell,n^\star) \right] \\
\bb (\ell,n^\star) &= \left[ b_1 (\ell,n^\star), \ldots, b_K (\ell,n^\star) \right]^\trp.
\end{align}
\end{subequations}
The single-rank updates in \eqref{update_a} can then be represented as
\begin{align}
\mA_{\ell+1}= \begin{bmatrix}
         \sqrt{\mu (\ell,n^\star) } \mA_{\ell} +\mD(\ell,n^\star)  \vspace*{2mm} \\
         \bb^\trp (\ell,n^\star)
         \end{bmatrix}.
\end{align}
\subsection{Maximum Ratio Transmission}
The \ac{mrt} precoder reads
\begin{align}
\mA_\ell=\beta_\mrt (\ell) \hspace*{.4mm} \mH_\ell^*
\end{align}
where $\beta_\mrt (\ell)=\left[\tr{\mH_\ell^* \mH_\ell^\trp}\right]^{-1/2}$. Consequently, $\beta_\mrt (\ell)$ is updated in each step by
\begin{align}
\frac{1}{\beta_\mrt^{2} (\ell+1)}=\frac{1}{\beta_\mrt^{2} (\ell)}+ \norm{\bg(n^\star)}^2. \label{eq:Beta_11}
\end{align}
From \eqref{eq:Beta_11}, one can show that
\begin{align}
\mA_{\ell+1}= \begin{bmatrix}
         \dfrac{\mA_\ell}{\sqrt{1+\beta_\mrt^2 (\ell) \norm{\bg(n^\star)}^2}}  \vspace*{2mm}\\
           \dfrac{\beta_\mrt (\ell) \hspace*{.4mm} \bg^\her(n^\star)}{\sqrt{1+\beta_\mrt^2 (\ell) \norm{\bg(n^\star)}^2}}  
         \end{bmatrix}.
\end{align}
As a result, it is concluded that $\mD(\ell,n^\star)=0$ and
\begin{subequations}
\begin{align}
\mu (\ell,n^\star) &= \left[ {1+\beta_\mrt^2 (\ell) \norm{\bg(n^\star)}^2}  \right]^{-1} \\
\bb (\ell,n^\star) &= \beta_\mrt (\ell) \sqrt{\mu (\ell,n^\star)} \hspace*{.4mm} \bg^* (n^\star) .
\end{align}
\end{subequations}
\subsection{Regularized Zero Forcing}
For \ac{rzf} precoding, we have
\begin{align}
\mA_\ell=\beta_\rzf(\ell;\lambda) \hspace*{.4mm} \mH^*_\ell \left(\mH_\ell^\trp \mH^*_\ell + \lambda \hspace*{.4mm} \mI_K \right)^{-1}
\end{align}
where the factor $\beta_\rzf(\ell;\lambda)$ reads
\begin{align}
\beta_\rzf (\ell;\lambda)=\left[\tr{\mJ_\ell(\lambda)^{-2} \mJ_\ell(0)}\right]^{-1/2}
\end{align}
with $\mJ_\ell(\lambda)\coloneqq\mH_\ell^\trp \mH^*_\ell + \lambda \hspace*{.4mm} \mI_K$. Noting that for any $\lambda$
\begin{align}
\mJ_{\ell+1}(\lambda)=\mJ_\ell(\lambda) + \bg(n^\star) \bg^\her(n^\star),
\end{align}
one utilizes the Sherman-Morrison formula \cite{bartlett1951inverse} and writes
\begin{align}
\mJ_{\ell+1}(\lambda)^{-1}=\mJ_\ell(\lambda)^{-1} -{ \br(\ell,n^\star;\lambda) \br^\her(\ell,n^\star;\lambda) },
\end{align}
where we define
\begin{align}
\br(\ell,n^\star;\lambda)\coloneqq \frac{\mJ_\ell(\lambda)^{-1} \bg(n^\star)}{\sqrt{1+\bg^\her(n^\star) \mJ_\ell(\lambda)^{-1} \bg(n^\star)}}
\end{align}
Note that $\bg^\her(n^\star) \mJ_\ell(\lambda)^{-1} \bg(n^\star)\geq 0$ for any $\lambda$, since $\mJ_\ell(\lambda)^{-1}$ is a positive semi-definite  matrix. Consequently, one can derive the rank-one update for $\beta_\rzf (\ell;\lambda)$ as
\begin{align}
\frac{1}{\beta_\rzf^{2} (\ell+1;\lambda)}=\frac{1}{\beta_\rzf^{2} (\ell;\lambda)} + \Delta (\ell,n^\star;\lambda) ,
\end{align}
where $\Delta (\ell,n^\star;\lambda)$ is defined in \eqref{eq:Delta_Big} on the top of the next page with $\mE(\ell,n^\star;\lambda)\in\setC^{K\times K}$ being defined as
\begin{align}
&\mE(\ell,n^\star;\lambda) \coloneqq \nonumber \\
&\left( \norm{\br (\ell,n^\star;\lambda)}^2 \mI_K  - \mJ_\ell(\lambda)^{-1}\right) \mJ_\ell(0) - \mJ_\ell(0) \mJ_\ell(\lambda)^{-1}.
\end{align}
Consequently, one can conclude that
\begin{subequations}
\begin{align}
\hspace*{-3mm} \mu (\ell,n^\star) &\hspace*{-1mm}=\hspace*{-1mm} \left[ {1+\beta_\rzf^{2} (\ell;\lambda) \Delta (\ell,n^\star;\lambda)}  \right]^{-1} \label{eq:aa}\\
\hspace*{-3mm} \mD (\ell,n^\star) &\hspace*{-1mm}=\hspace*{-1mm} \beta_\rzf (\ell;\lambda) \sqrt{\mu (\ell,n^\star)} \mH^*_\ell \br (\ell,n^\star;\lambda) \br^\her (\ell,n^\star;\lambda) \\
\hspace*{-3mm} \bb (\ell,n^\star) &\hspace*{-1mm}=\hspace*{-1mm} \sqrt{\dfrac{\beta^2_\rzf (\ell;\lambda) \mu (\ell,n^\star)}{ 1+\bg^\her(n^\star) \mJ_\ell(\lambda)^{-1} \bg(n^\star) }} \hspace*{1mm} \br^* (\ell,n^\star;\lambda)\label{eq:cc}
\end{align}
\end{subequations}
where we further drop $\lambda$ in the left hand sides of \eqref{eq:aa}-\eqref{eq:cc} for the sake of compactness.
%
\subsection{Zero Forcing}
The \ac{zf} precoder deduces from \ac{rzf} precoding by setting $\lambda=0$. In this case,
\begin{align}
\mA_\ell=\beta_\zf (\ell) \hspace*{.4mm} \mH_\ell^*  \mJ_{\ell}(0)^{-1}
\end{align}
in which $\beta_\zf (\ell)=\beta_\rzf (\ell;0)$. By setting $\lambda=0$ in the results derived for \ac{rzf}, we have
\begin{align}
\frac{1}{\beta_\zf^{2} (\ell+1)}=\frac{1}{\beta_\rzf^{2} (\ell)} - \norm{\br_0 (\ell,n^\star)}^2 ,
\end{align}
where we define $\br_0 (\ell,n^\star) \coloneqq \br (\ell,n^\star;0)$. Moreover, 
\begin{subequations}
\begin{align}
\hspace*{-2mm} \mu (\ell,n^\star) &= \left[ {1-\beta_\zf^{2} (\ell) \norm{\br_0 (\ell,n^\star)}^2}  \right]^{-1} \\
\hspace*{-2mm} \mD (\ell,n^\star) &= \beta_\zf (\ell) \sqrt{\mu (\ell,n^\star)} \hspace*{1mm} \mH^*_\ell \br_0 (\ell,n^\star) \br_0^\her (\ell,n^\star) \\
\hspace*{-2mm} \bb (\ell,n^\star) &= \sqrt{\dfrac{\beta^2_\zf (\ell) \mu (\ell,n^\star)}{ 1+\bg^\her(n^\star) \mJ_\ell(0)^{-1} \bg(n^\star) }} \hspace*{1mm} \br_0^* (\ell,n^\star).
\end{align}
\end{subequations}
%
%

\section{Numerical Results}
Throughout the numerical investigations, we have considered \ac{iid} Rayleigh fading channels. Hence, the entries of $\mG$ are \ac{iid} zero-mean and unit-variance complex Gaussian random variables. The system parameters, such as the energy~efficiency~and~number of selected antennas, are then averaged over multiple realizations numerically. Due to the linear computational complexity of the proposed algorithm, Monte-Carlo simulations are feasible even for large dimensions. It is worth moreover to indicate that the optimal search for the dimensions considered here are practically infeasible due to its high computational complexity.

\begin{figure}[t]
\hspace*{-1.3cm}  
\resizebox{1.25\linewidth}{!}{

\pstool[width=.35\linewidth]{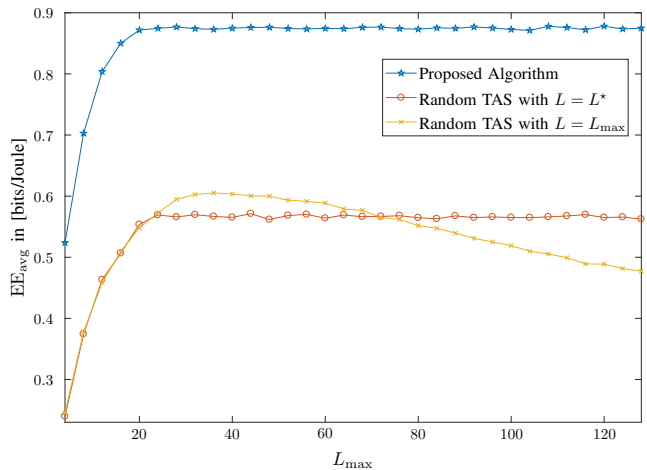}{

\psfrag{EE}[c][c][0.22]{$\mathrm{EE}_{\avg}$ in [bits/Joule]}
\psfrag{L-max}[c][t][0.22]{$L_{\max}$}
\psfrag{AlgorithmP-opt-L-opt-XXX111A}[l][l][0.2]{Proposed Algorithm}
\psfrag{AlgorithmP-opt-L-rnd-XXX111A}[l][l][0.2]{Random TAS with $L=L^\star$}
\psfrag{AlgorithmP-opt-L-rnd-FXX111A}[l][l][0.2]{Random TAS with $L=L_{\max}$ }

\psfrag{20}[c][c][0.18]{$20$}
\psfrag{40}[c][c][0.18]{$40$}
\psfrag{60}[c][c][0.18]{$60$}
\psfrag{80}[c][c][0.18]{$80$}
\psfrag{100}[c][c][0.18]{$100$}
\psfrag{120}[c][c][0.18]{$120$}

\psfrag{0.3}[r][c][0.18]{$0.3$}
\psfrag{0.4}[r][c][0.18]{$0.4$}
\psfrag{0.5}[r][c][0.18]{$0.5$}
\psfrag{0.6}[r][c][0.18]{$0.6$}
\psfrag{0.7}[r][c][0.18]{$0.7$}
\psfrag{0.8}[r][c][0.18]{$0.8$}
\psfrag{0.9}[r][c][0.18]{$0.9$}

}
}
\caption{Energy efficiency versus the maximum number of selected antennas $L_{\max}$ considering $N=128$ transmit antennas and $K=4$ users. Here, the power constraint is $P_{\max}=0$~dB and the parameters of \ac{rf}-chains are $\xi^{-1}=0.4$, $Q_{\cir}^{\rm Rx} =Q_{\cir}^{\rm Tx} = 48$ mW and $Q_{\sync}=62$ mW. As the figure shows, the proposed algorithm significantly outperforms the random \ac{tas} approaches with power control.}
\label{fig:1}
\end{figure}
For the \ac{mrt} precoder, we have sketched the average energy efficiency defined as
\begin{align}
\mathrm{EE}_{\avg}= \E \set{ \mae(L^\star,P^\star|\bw) }
\end{align}
versus $L_{\max}$ where $L^\star$ and $P^\star$ are the outputs of Algorithm~\ref{alg1} for the \ac{mrt} precoder and $\mae(\ell,P|\bw)$. The performance of the proposed algorithm is moreover compared to the performance of random \ac{tas} algorithms with power control. 

Fig.~\ref{fig:1} shows the energy efficiency as a function of $L_{\max}$ for $K=4$ users and $N=128$ available transmit antennas when $K\leq L_{\max} \leq N$. The transmit power is considered to be limited by $P_{\max}=0$~dB and the transmit and receive power amplifiers are assumed to have efficiency factor $\xi^{-1}=0.4$. The consumed power at the \ac{rf} chains and synchronization oscillators are moreover set to be $Q^{\rm Tx}_{\cir} = Q^{\rm Rx}_{\cir} = 48$ mW and $Q_{\sync}=62$ mW. In this figure, the blue line shows the performance of the proposed algorithm in which the selection subset as well as the transmit power is being iteratively tuned via the stepwise approach. For the sake of comparison, we have also plotted the energy efficiency of two random \ac{tas} algorithms with power control. The first random algorithm, whose performance is depicted by the yellow line, selects $L_{\max}$ out of $N$ transmit antennas and chooses the transmit power such that the energy efficiency is maximized. In the other random approach, shown in the figure with the red line, the number of selected transmit antennas for each $L_{\max}$ is set to $L^\star$ in which $L^\star$ denotes the number of selected antennas via the proposed algorithm and is a function of $L_{\max}$. The latter random algorithm then selects $L^\star$ transmit antennas at random and optimize the energy efficiency with respect to $P$ subject to $P\leq P_{\max}$. As Fig.~\ref{fig:1} depicts, the proposed algorithm significantly outperforms the random \ac{tas} approaches with power control\footnote{Note that when $L_{\max}$ grows with $N$, random selection is a good bound on the performance of several practical \ac{tas} algorithms; see \cite{bereyhi2017nonlinear}.}. The figure furthermore shows that even in the case of random selection, the increase in the number of selected antennas does not necessarily improve the performance. In fact for large values of $L_{\max}$, the random algorithm with $L^\star < L_{\max}$ selected antennas enhances the energy efficiency compared to the random \ac{tas} approach with $L_{\max}$ active transmit antennas.

\begin{figure}[t]
\hspace*{-1.3cm}  
\resizebox{1.25\linewidth}{!}{

\pstool[width=.35\linewidth]{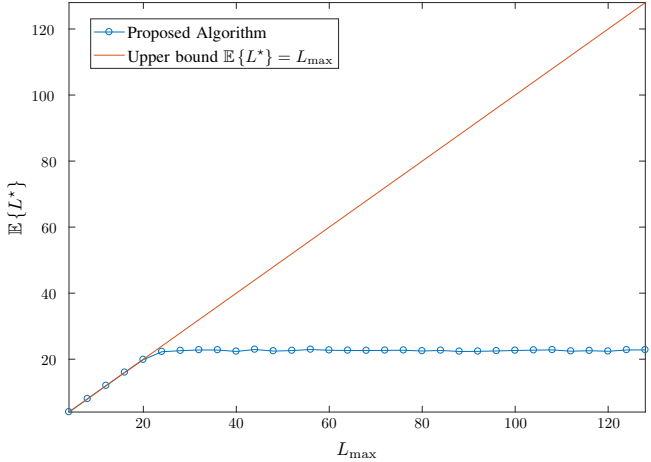}{

\psfrag{L-star}[c][c][0.22]{$\E\{L^\star \}$}
\psfrag{L-max}[c][t][0.22]{$L_{\max}$}
\psfrag{Algorithm-Lav-Lmax-AADDDAA}[l][l][0.2]{Proposed Algorithm}
\psfrag{Algorithm-Lav-Lmax-BBDDDAA}[l][l][0.2]{Upper bound $\E\{L^\star\}=L_{\max}$}

\psfrag{20}[c][c][0.18]{$20$}
\psfrag{40}[c][c][0.18]{$40$}
\psfrag{60}[c][c][0.18]{$60$}
\psfrag{80}[c][c][0.18]{$80$}
\psfrag{100}[c][c][0.18]{$100$}
\psfrag{120}[c][c][0.18]{$120$}

\psfrag{20a}[r][c][0.18]{$20$}
\psfrag{40a}[r][c][0.18]{$40$}
\psfrag{60a}[r][c][0.18]{$60$}
\psfrag{80a}[r][c][0.18]{$80$}
\psfrag{100a}[r][c][0.18]{$100$}
\psfrag{120a}[r][c][0.18]{$120$}

}
}
\caption{Average number of active antennas selected via the proposed algorithm versus the maximum number of selected antennas considering $N=128$ and $K=4$. Here, it is assumed that $P_{\max}=0$~dB, $\xi^{-1}=0.4$, $Q_{\cir}^{\rm Rx} =Q_{\cir}^{\rm Tx} = 48$ mW and $Q_{\sync}=62$ mW. The figure depicts that the algorithm stops selecting transmit antennas at $L^\star \approx 24$ which is significantly smaller than $N$.}
\label{fig:2}
\end{figure}
\begin{figure}[t]
\hspace*{-1.3cm}  
\resizebox{1.25\linewidth}{!}{

\pstool[width=.35\linewidth]{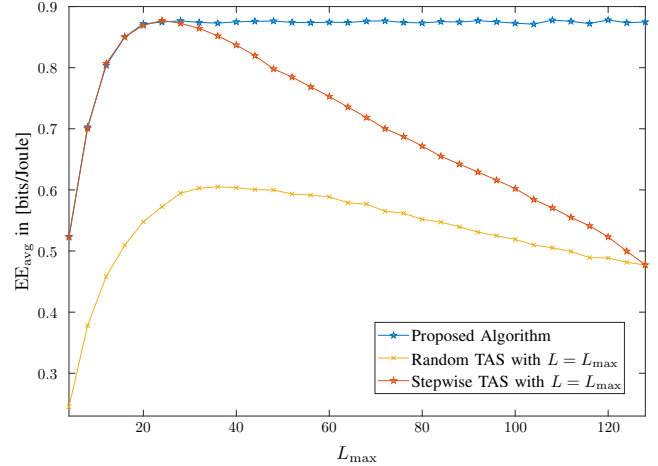}{

\psfrag{EE}[c][c][0.22]{$\mathrm{EE}_{\avg}$ in [bits/Joule]}
\psfrag{L-max}[c][t][0.22]{$L_{\max}$}
\psfrag{AlgorithmP-opt-L-opt-XAAABBB}[l][l][0.2]{Proposed Algorithm}
\psfrag{AlgorithmP-opt-L-opt-TAAABBB}[l][l][0.2]{Stepwise TAS with $L=L_{\max}$}
\psfrag{AlgorithmP-opt-L-rnd-FAAABBB}[l][l][0.2]{Random TAS with $L=L_{\max}$ }

\psfrag{20}[c][c][0.18]{$20$}
\psfrag{40}[c][c][0.18]{$40$}
\psfrag{60}[c][c][0.18]{$60$}
\psfrag{80}[c][c][0.18]{$80$}
\psfrag{100}[c][c][0.18]{$100$}
\psfrag{120}[c][c][0.18]{$120$}

\psfrag{0.3}[r][c][0.18]{$0.3$}
\psfrag{0.4}[r][c][0.18]{$0.4$}
\psfrag{0.5}[r][c][0.18]{$0.5$}
\psfrag{0.6}[r][c][0.18]{$0.6$}
\psfrag{0.7}[r][c][0.18]{$0.7$}
\psfrag{0.8}[r][c][0.18]{$0.8$}
\psfrag{0.9}[r][c][0.18]{$0.9$}

}
}
\caption{Energy efficiency versus the maximum number of selected antennas considering the proposed algorithm as well as the modified version of the stepwise approach in which exactly $L_{\max}$ antennas are selected. The parameters are set similar to those in Fig.~\ref{fig:1}. As it depicts, the energy efficiency in the modified algorithm is maximized at $L_{\max} = 24$ which agrees with the results reported in Fig.~\ref{fig:2}.}
\label{fig:3}
\end{figure}

In order to illustrate the latter observation, we have further sketched in Fig.~\ref{fig:2} the average number of selected antennas given by the proposed algorithm, i.e., $\E \{L^\star\}$, versus $L_{\max}$ for the setting considered in Fig.~\ref{fig:1}. As it demonstrates, the algorithm up to $L_{\max}=23$ selects the almost $L_{\max}$ transmit antennas. Nevertheless, for larger values of $L_{\max}$ it stops to select more antennas around $L^\star \approx 24$. This means that further growth in the number of selected antennas degrades the energy efficiency of the system. Such an observation is intuitive, due to the fact that the growth in the number of active antennas can both increase the rate and consumed power. In order to observe the latter intuition further, we have plotted the energy efficiency of a modified version of the proposed algorithm, for the setting considered in previous figures, in Fig.~\ref{fig:3}. In this version of the algorithm, the proposed algorithm is enforced to select exactly $L_{\max}$ transmit antennas via the stepwise regression method. In fact in this case, the algorithm does not break at $L^\star$ and uses forward selection to collect a subset of $L_{\max}$ transmit antennas. As the figure shows, the modified algorithm meets the maximum around $L_{\max}\approx 24$ which agrees with the observation in Fig.~\ref{fig:2}. At $L_{\max}=N$, moreover, it lies on the random \ac{tas} since they both select all transmit antennas and optimize the transmit power subject to the energy efficiency. Noting that the proposed algorithm needs $L^\star$ number of iterations, the result indicates that the complexity is further reduced for $L_{\max}$ growing with $N$, since in the large limit $L^\star$ does not necessarily grow with $L_{\max}$.

%
%
%
%
\section{Conclusion}
\label{conclusion}
This paper has proposed a class of iterative algorithms for \ac{tas} and transmit power control in downlink massive multiuser \ac{mimo} systems with linear precoders. The algorithms follow the approach of stepwise regression methods and enjoy low computational complexity. The results have shown significant gains obtained via the proposed algorithm. It has been also demonstrated that antenna selection, in addition to \ac{rf}-cost reduction, could enhance the performance in several massive \ac{mimo} settings. The derivations given in this paper can be more enlightened by further numerical investigations which will be added in the forthcoming version of the manuscript. Using tools from large-system analysis, the asymptotic outputs of the algorithm can be derived in terms of explicit fixed-point equations. Such investigations are possible extensions of this work and are currently ongoing. 

\bibliography{ref}
\bibliographystyle{IEEEtran}

\begin{acronym}
\acro{mimo}[MIMO]{Multiple-Input Multiple-Output}
\acro{csi}[CSI]{Channel State Information}
\acro{awgn}[AWGN]{Additive White Gaussian Noise}
\acro{iid}[i.i.d.]{independent and identically distributed}
\acro{ut}[UT]{User Terminal}
\acro{bs}[BS]{Base Station}
\acro{tas}[TAS]{Transmit Antenna Selection}
\acro{lse}[LSE]{Least Squared Error}
\acro{rhs}[r.h.s.]{right hand side}
\acro{lhs}[l.h.s.]{left hand side}
\acro{wrt}[w.r.t.]{with respect to}
\acro{rs}[RS]{Replica Symmetry}
\acro{rsb}[RSB]{Replica Symmetry Breaking}
\acro{papr}[PAPR]{Peak-to-Average Power Ratio}
\acro{mrt}[MRT]{Maximum Ratio Transmission}
\acro{zf}[ZF]{Zero Forcing}
\acro{rzf}[RZF]{Regularized Zero Forcing}
\acro{snr}[SNR]{Signal-to-Noise Ratio}
\acro{rf}[RF]{Radio Frequency}
\acro{mf}[MF]{Match Filtering}
\acro{tdd}[TDD]{Time-Division Duplexing}
\end{acronym}

\end{document}